\documentclass[a4paper,10pt,twoside]{cpc-hepnp}

\usepackage{CJK}
\usepackage{multicol}
\usepackage{graphicx}
\usepackage{epsfig}
\usepackage{booktabs}
\usepackage{amssymb,bm,mathrsfs,bbm,amscd}
\usepackage[tbtags]{amsmath}
\usepackage{lastpage}

\newcommand{\rr} {\boldsymbol{r}}

\allowdisplaybreaks

\begin{document}
\begin{CJK*}{GBK}{song}

\fancyhead[c]{\small Chinese Physics C~~~Vol. XX, No. X (201X)
XXXXXX} \fancyfoot[C]{\small 010201-\thepage}

\footnotetext[0]{Received 31 March 2015}

\title{Study of weakly-bound odd-A nuclei with quasiparticle blocking\thanks{Supported by National Key Basic Research Program of China (2013CB83440), National Natural Science Foundation of China (11375016, 11235001 and 11320101004) and Research Fund for the Doctoral Program of Higher Education of China (20130001110001) }}

\author{%
     XIONG Xue-Yu$^{1}$
\quad PEI Jun-Chen$^{1;1)}$\email{peij@pku.edu.cn}%
\quad ZHANG Yi-Nu$^{1}$
\quad ZHU Yi$^{1}$
}

\maketitle

\address{%
$^1$ State Key Laboratory of Nuclear Physics and Technology, School of Physics, Peking University, Beijing 100871, China
}

\begin{abstract}
The coordinate-space Hartree-Fock-Bogoliubov (HFB) approach with quasiparticle blocking has been applied to study the odd-A weakly bound nuclei $^{17,19}$B and $^{37}$Mg, in which
halo structures have been reported in experiments.  The Skyrme nuclear forces SLy4 and UNEDF1 have been adopted in our calculations. The results with and without blocking have been compared to demonstrate the emergence of deformed halo structures due to blocking effects. In our calculations, $^{19}$B and $^{37}$Mg have remarkable features of deformed halos.
\end{abstract}



\footnotetext[0]{\hspace*{-3mm}\raisebox{0.3ex}{$\scriptstyle\copyright$}2015
Chinese Physical Society and the Institute of High Energy Physics
of the Chinese Academy of Sciences and the Institute
of Modern Physics of the Chinese Academy of Sciences and IOP Publishing Ltd}%

\begin{multicols}{2}

\section{Introduction}
Nuclei with extremely unbalanced neutron to proton ratios, close to drip lines,  can have a lot of exotic phenomena such as halo states and pygmy resonances\cite{halo-exp}. From the theoretical point of view, drip-line nuclei are weakly bound superfluid quantum many-body systems and can offer a unique environment to test threshold effects as well as to investigate effective  interactions.
In the experimental aspect, the new-generation radioactive-isotope  beam facilities such as FRIB \cite{FRIB} and HIAF ~\cite{HIAF} are being built to explore new physics near drip lines and to study nuclear processes in astrophysical environments.

One of the most important reasons to study extremely neutron-rich nuclei is their exotic surface properties and dynamics, due to the intertwining of the halo decoupling effect, surface deformations, pairing correlations and continuum couplings\cite{Pei13,Pei14a,RBF1}. In particular, the deformed halo structure can have decoupled core and halo deformations and various situations can happen~\cite{sgzhou}. In a series of previous studies based on the coordinate-space Hartree-Fock-Bogoliubov approach, we have studied exotic deformed halos and collective modes in even-even weakly bound nuclei\cite{Pei13,Pei14a}. Actually deformed halos in experiments up to now are all odd-mass nuclei such as $^{11}$Be~\cite{Be11}, $^{31}$Ne~\cite{Ne31} and $^{37}$Mg~\cite{Mg37}. Theoretical studies of the odd-A nuclei is not so popular  because of additional computational complexity  when dealing with pairing correlations. Indeed, the quasiparticle blocking effect has to be taken into account for spin-polarized systems such
as nuclei and cold atomic gases~\cite{Bertsch}.

In the past few years, several odd-N halo nuclei with deformations have been reported experimentally, such as $^{31}$Ne~\cite{Ne31} and $^{37}$Mg~\cite{Mg37}, implying more opportunities than expected.
Deformations and weak-binding effects together can change the standard Nilsson shell evolution and drive $s$ or $p$ orbits that are not even around Fermi surfaces into halo states.
Very recently, the deformed halo $^{37}$Mg was observed experimentally and currently $^{37}$Mg is the heaviest halo nucleus\cite{Mg37}. A large $p$-wave component is required to reproduce the experiment.  Theoretical studies are few since the experiment result of $^{37}$Mg has come out only very recently. In addition to the odd-N halo nuclei, we are also interested in the odd-Z halo nuclei such as $^{17}$B and $^{19}$B\cite{Suzuki99}. In fact the neutron halo could also be influenced by the polarization effect due to the blocking of
proton orbits. Several experimental and theoretical studies suggest $^{17}$B has a halo structure\cite{Suzuki02,Ren,HU08,xiajijuan12,Guoyanqing10,WangM08}.  For $^{19}$B, the experimental studies suggest it may have a $4n$-halo structure\cite{Suzuki99}. Recent experiments found the valence neutron of $^{19}$B has large $d$-wave component which hinders the formation of the halo structure\cite{Gaudefroy12}.   Therefore the purpose of this work is to employ the Hartree-Fock-Bogoliubov(HFB) approach with quasiparticle blocking in deformed coordinate space to study these odd-A halo nuclei.

HFB theory can properly treat the coupling between the weak-bound orbits and continuum spectrum, which is essential for weakly bound nuclei. The continuum states affect not only the properties of ground states but also excited states~\cite{Pei14a,Mazy}.  Solving the HFB equation in coordinate-space can properly and self-consistently taken into account
the deformations, pairing, and weak-binding effects~\cite{Pei13}. In this work, the Skyrme-HFB equation is solved in deformed coordinate spaces by the HFB-AX solver \cite{Pei08}, where B-spline techniques are adopted to treat axial symmetric deformed nuclei within a two-dimensional (2D) lattice box. The hybrid MPI+OpenMP
parallel programming is performed to obtain converged results in a reasonable time~\cite{Pei13}. Note that 3D coordinate-space HFB that uses multi-wavelet analysis has recently been developed\cite{Pei14}. For the particle-hole interaction channel, the UNEDF1 and SLy4 parameterizations of the Skyrme force are adopted. The SLy4 force has been obtained
by constraining the properties of neutron matter and has been widely used for neutron-rich nuclei~\cite{sly4}. The latest UNEDF1 force is obtained by massively fitting global nuclear properties and is very precise for the whole nuclear landscape~\cite{unedf1}.
For the particle-particle channel, the density-dependent surface pairing interactions is employed. The pairing strength is taken as $V_0$=500 MeV for SLy4 and 460 MeV for UNEDF1 to reproduce reasonable neutron pairing gaps in $^{120}$Sn. Note that the surface pairing has also been used in relativistic Hartree-Bogoliubov calculations for halo nuclei~\cite{cheny}.

This paper is organized as  follows. In Sec.2 the formalism of the quasiparticle blocking method is introduced. In Sec.3 numerical calculations and discussions of halo nuclei $^{17}$B, $^{19}$B, $^{37}$Mg are presented. Finally, the main conclusion of this paper and perspectives are given in the summary.

\section{Theoretical framework }
The HFB equation in the coordinate-space representation can be written as~\cite{Pei08}:

\begin{equation}\label{1}
\left[
  \begin{array}{cc}
    h(\rr)-\lambda & \Delta(\rr) \\
    \Delta^{*}(\rr) & -h(\rr)+\lambda \\
  \end{array}
\right]
  \left[
    \begin{array}{c}
      U_{k}(\rr) \\
      V_{k}(\rr) \\
    \end{array}
  \right]
  = E_k
    \left[
    \begin{array}{c}
      U_{k}(\rr) \\
      V_{k}(\rr) \\
    \end{array}
  \right],
\end{equation}
where $h$ is the Hartree-Fock Hamiltonian; $\Delta$ is the pairing potential;
$U_k$ and $V_k$ are the upper and lower components of quasi-particle
wave functions, respectively; $E_k$ is the quasi-particle energy; and
$\lambda$ is the Fermi energy (or chemical potential).
For bound systems, $\lambda < 0$ and the self-consistent densities and fields
are localized in space.
For $|E_k|<-\lambda$, the eigenstates are discrete and
$V_k(\rr)$ and $U_k(\rr)$ decay exponentially.
The quasiparticle continuum corresponds to  $|E_k|>-\lambda$. For those states, the upper component of the
wave function always has  a scattering asymptotic form. By applying the
box boundary condition, the continuum becomes discretized and one obtains
a finite number of continuum quasi-particles. In principle, the box solution
representing the continuum can be  close to the exact
solution when a sufficiently big box and small mesh size are adopted.
In this work, the  maximum mesh spacing is 0.6\,fm, the box is 27 fm and the order of the B-splines is 12.

In our approach, the systems are assumed to have axial and reflection symmetries.
Based on the quasiparticle wave functions, the particle density $\rho(\mathbf{r})$ and the pairing density $\tilde{\rho}(\mathbf{r})$ of even-even nuclei  can be written as
\begin{equation}
\begin{array}{c}
\rho(\mathbf{r})=2\sum_k V_k^{*}(\rr)V_k(\rr), \vspace{3pt}\\
\tilde{\rho}(\mathbf{r})=-2\sum_k V_k(\rr)U_k^{*}(\rr),
\end{array}
\end{equation}
where in the sum the quasiparticle energy cutoff is taken as ($60-\lambda$) MeV. With the quasiparticle blocking of the state $\mu$, the density $\rho_{B}^{~\mu}(\mathbf{r})$ and the pairing density $\tilde{\rho}_{B}^{~\mu}(\mathbf{r})$ of odd-mass nuclei in the coordinate space can be
written as~\cite{Bertsch}:

\begin{eqnarray}
 \rho_{B}^{~\mu}(\mathbf{r})&=& U_{\mu}(\rr)U_{\mu}^{*}(\rr)+V_{\mu}^{*}(\rr)V_{\mu}(\rr) +2\sum_{k\neq \mu} V_k^{*}(\rr)V_k(\rr),\nonumber\\
 \tilde{\rho}_{B}^{~\mu}(\mathbf{r})&=&-2\sum_{k\neq \mu} V_k(\rr)U_k^{*}(\rr).
\end{eqnarray}
Similarly, the blocking method has also been discussed extensively in the basis expansion method~\cite{Xu98,Cwiok99,RHB2}. In addition to nuclear systems, the blocking method
has also been widely used in studies of spin-polarized cold atomic gases~\cite{Pei10,2FLA}.
  The particle number equation for the odd-A system has to be modified as:
\begin{equation}
N=1+2\Sigma_{k\neq \mu}v_k^2=\int d\rr \rho_{B}^{~\mu}(\rr).
\end{equation}

\end{multicols}

\begin{figure}[htbp]
\begin{center} \scalebox{0.58}{\includegraphics{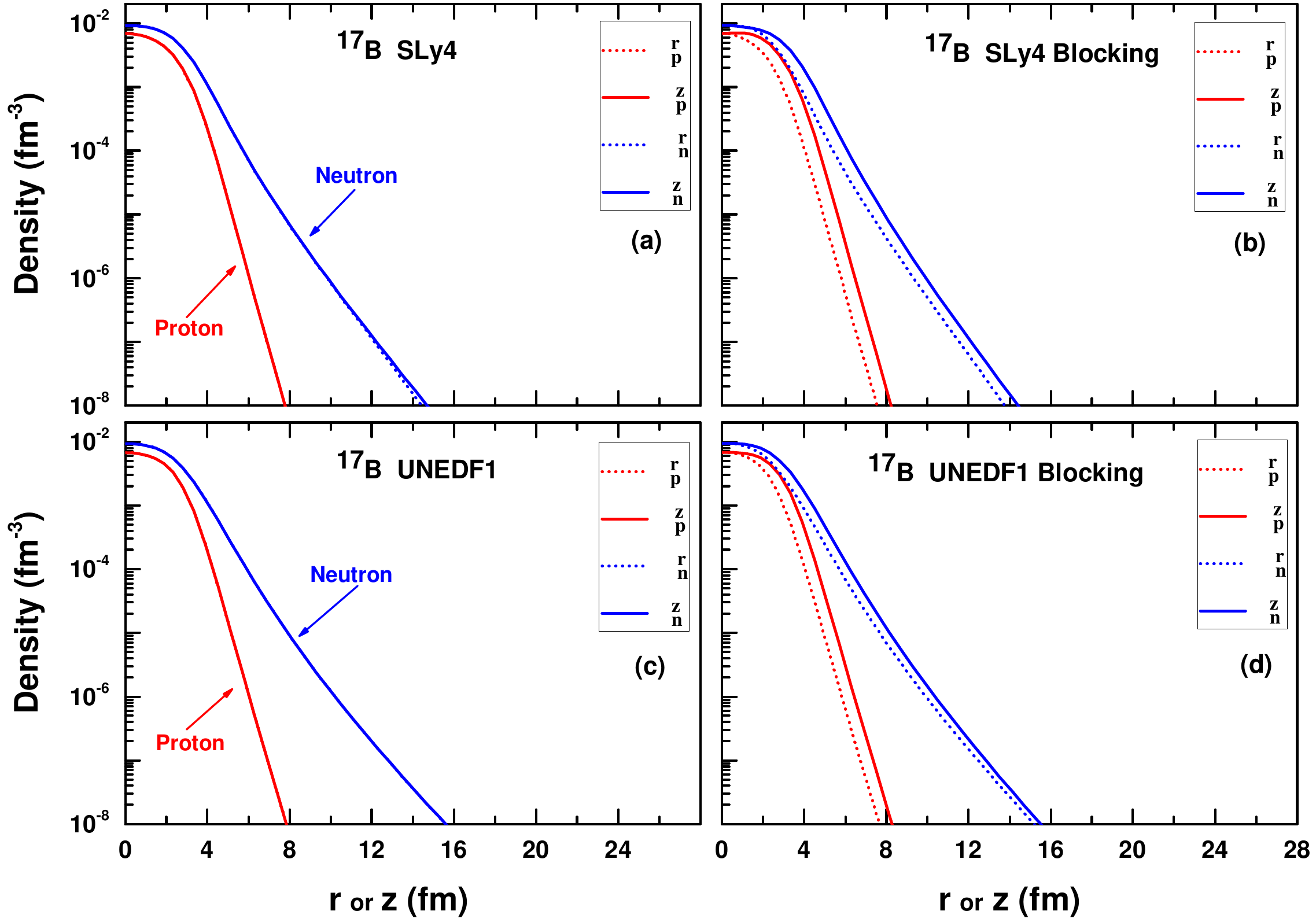}}
\end{center}
\caption{\label{figB17} (Color online) The density distributions of $^{17}$B which are obtained by calculations with (a) the SLy4 force without blocking, (b) the SLy4 force with blocking, (c) the UNEDF1 force without blocking, (d) the UNEDF1 force with blocking. The dashed lines represent the density profiles along the cylindrical coordinates $r$-axis $\rho(r,z=0)$ and solid lines represent density profiles along the $z$-axis $\rho(r=0,z)$, respectively.}
\end{figure}

\begin{figure}[htbp]
\begin{center}
 \scalebox{0.58}{\includegraphics{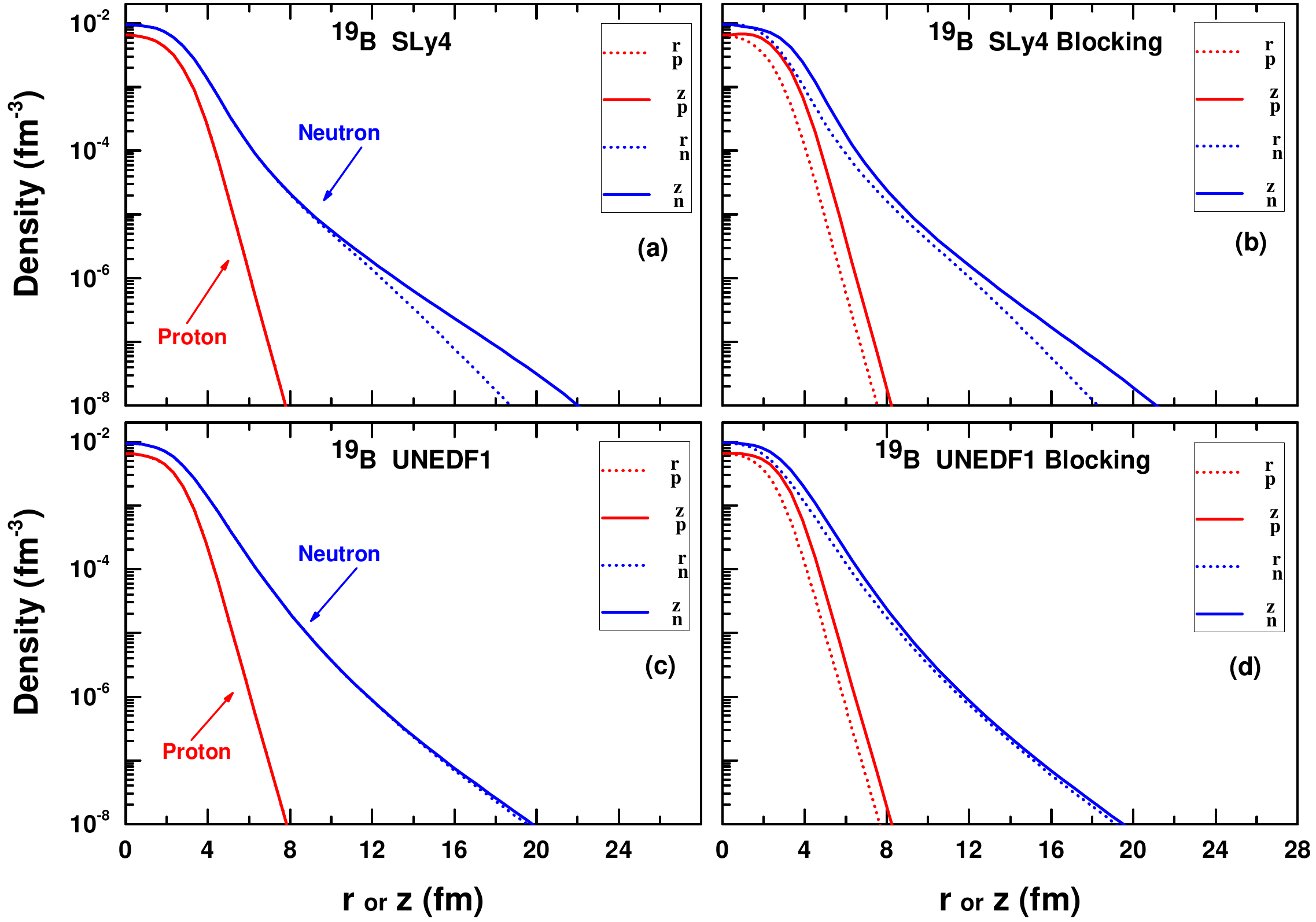}}
 \end{center}
\caption{\label{figB19} (Color online) Similar to Fig. 1 but for density distributions of $^{19}$B.}
\end{figure}

\begin{figure}[htbp]
\begin{center}
 \scalebox{0.58}{\includegraphics{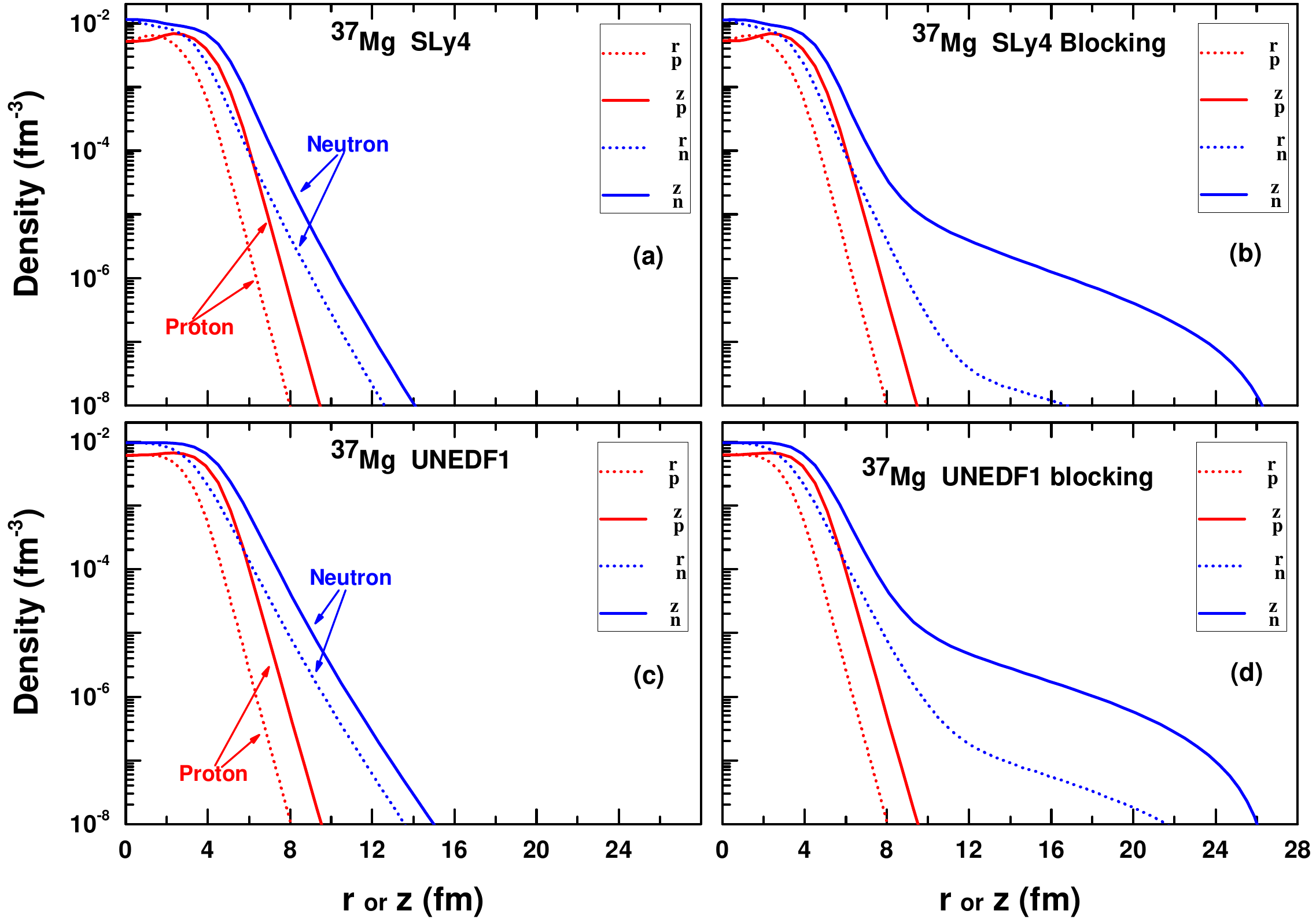}}
  \end{center}
\caption{\label{figMg37} (Color online) Similar to Fig.1 but for density distributions of $^{37}$Mg.}
\end{figure}

\begin{figure}[htbp]
\begin{center}
 \scalebox{0.30}{\includegraphics{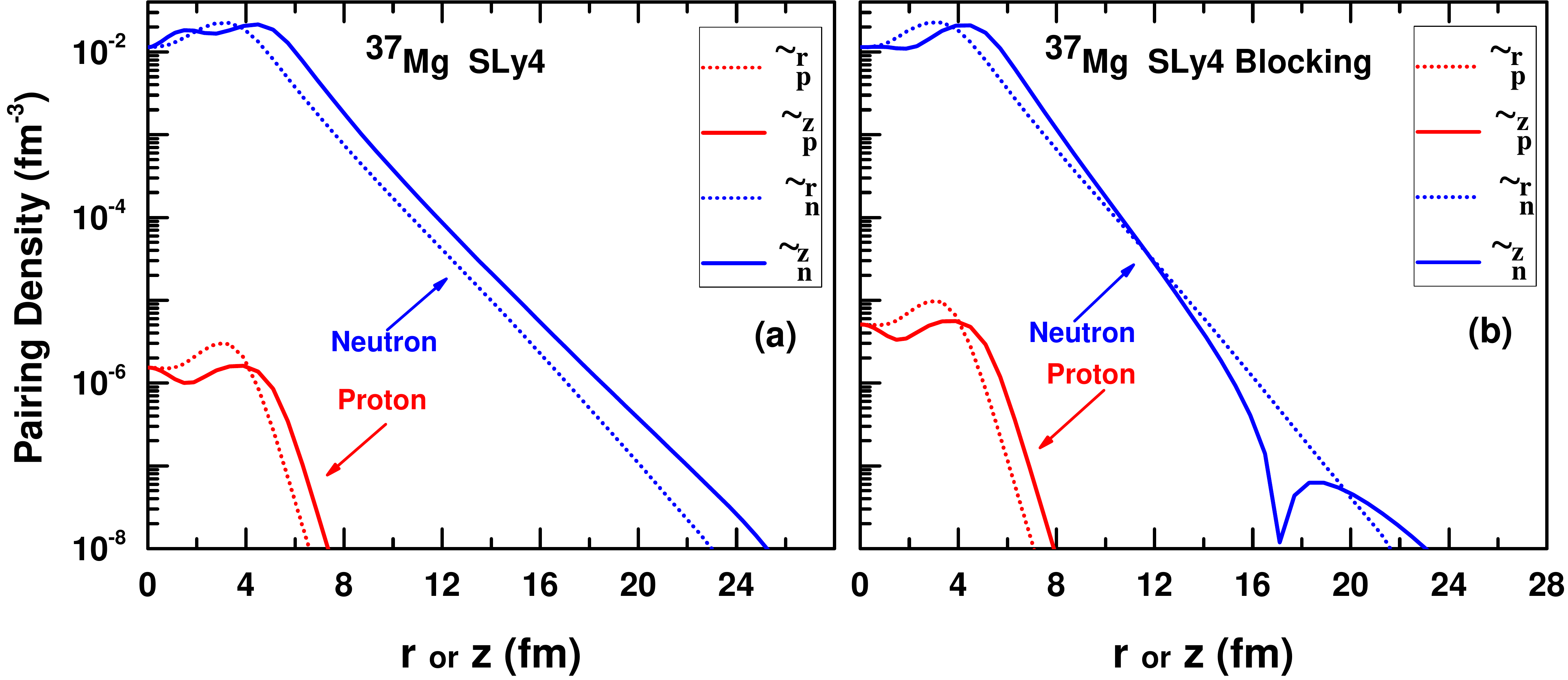}}
 \end{center}
\caption{ \label{figMg37p} (Color online) Similar to Fig.1 but for pairing density distributions of $^{37}$Mg obtained
by calculations with the SLy4 force. Note that absolute values of negative pairing densities are plotted.   }
\end{figure}

\begin{table}[h]
\begin{center}
\caption{\label{table1}
Calculation results without or with blocking for binding energies $E^{wob}$ and $E^{wb}$, Fermi energies $\lambda^{wob}_{p,n}$ and $\lambda^{wb}_{p,n}$,    quadrupole deformations $\beta_2^{wob}$ and $\beta_{2}^{wb}$, r.m.s. radii $r_{p,n}^{wob}$ and $r_{p,n}^{wb}$, pairing energies $E_{pair}^{wob}$ and $E_{pair}^{wb}$, pairing gaps $\Delta_{p,n}^{wob}$ and $\Delta_{p,n}^{wb}$ of nuclei $^{17}$B, $^{19}$B and $^{37}$Mg with the SLy4 and UNEDF1 forces. The subscripts $wb$ and $wob$  mean the calculations are carried out with or without blocking. Energies are given in units of MeV and radii are given in units of fm.}
\begin{tabular}{|c|ccc|ccc|}
\hline
&~&SLy4&~&~&UNEDF1&~\\
\hline
 &$^{17}$B &$^{19}$B &$^{37}$Mg &$^{17}$B &$^{19}$B &$^{37}$Mg\\
\hline
$E^{wob}$& -98.69 & -98.81 & -266.14 & -89.42 & -90.31 & -259.56 \\
$E^{wb}$ &-96.97 & -97.20 & -263.97&-87.81 & -88.80 & -257.62 \\
$S_{2n}^{wob}$&1.80  & 0.12 & $S_{n}(1.82)$ & 2.29 & 0.90 & $S_{n}(2.37)$\\
$S_{2n}^{wb}$ & 2.13&0.23  & $S_{n}(-0.58)$ & 2.44& 0.98 & $S_{n}(0.43)$ \\
$\lambda^{wob}_{p}$ & -23.31 & -25.32 & -22.22  & -22.37 & -24.16 & -21.46\\
$\lambda^{wb}_{p}$ &-22.28 & -21.79 & -22.16 &-21.66 & -23.49 & -21.17\\
$\lambda^{wob}_{n}$ & -1.32 & -0.56&-1.98 & -0.73 & -0.11&-1.63\\
$\lambda^{wb}_{n}$ &-1.44 & -0.59&-2.41 &-0.79 & -0.14&-1.81\\
$\beta_2^{wob}$ & 0.001 & 0.03& 0.39 & 0.001 & 0.001& 0.41\\
$\beta_2^{wb}$ &0.30& 0.31 & 0.45 &0.24& 0.21 & 0.49\\
$r_{p}^{wob}$ &2.55  &2.58 &3.19  &2.52  &2.55 &3.19 \\
$r_{p}^{wb}$ &2.55 &2.59  & 3.18 & 2.53& 2.56 &3.18 \\
$r_{n}^{wob}$ & 3.27 & 3.71&3.60  &3.44  & 3.80& 3.70\\
$r_{n}^{wb}$ &3.25 &3.68  &3.73  &3.41 & 3.78 &4.07 \\
$r_{m}^{wob}$ &3.09  &3.45 &3.52  &3.20  &3.51 &3.58 \\
$r_{m}^{wb}$ &3.06 &3.43  &3.62  &3.18 &3.50  &3.84 \\
$E_{pair}^{wob}$ & -36.08 & -42.50&-20.81  & -31.24 &-37.04 & -30.55\\
$E_{pair}^{wb}$ &-28.16 &-36.07  & -13.47 & -25.42& -31.62 &-24.98 \\
$\Delta_{p}^{wob}$ &2.03  &1.89 &0.20  & 1.93 & 1.84& 0.01\\
$\Delta_{p}^{wb}$ &$10^{-5}$ &$10^{-6}$  &$10^{-5}$ &$10^{-4}$ &$10^{-5}$  & $10^{-4}$\\
$\Delta_{n}^{wob}$ &3.76  &3.72 & 2.10 &3.42  & 3.49& 2.48\\
$\Delta_{n}^{wb}$ & 3.61& 3.62 &1.74 &3.34 &3.42  &2.24 \\
\hline
\end{tabular}
\end{center}
\end{table}
\begin{multicols}{2}
\section{Results and discussion}
We have studied the ground state properties of $^{17,19}$B and $^{37}$Mg with and without  quasiparticle blocking.  The density distributions,
deformations, and energetic properties are presented in this section.  To check the dependence on Skyrme forces, we adopted the usually used SLy4 force
and the newly developed UNEDF1 force.

Fig.\ref{figB17} displays the density profiles of $^{17}$B calculated with and without blocking method with SLy4 force and UNEDF1 force.
The densities are displayed along the cylindrical  coordinates $z$-axis (the axis of symmetry)  and $r$-axis (the axis perpendicular to $z$-axis and $r$=$\sqrt{x^2+y^2}$), respectively. The differences between the density profiles $\rho_{z({r=0})}$ and  $\rho_{r({z=0})}$  actually reflect the surface deformations.
 The highest state of proton $1p_{1/2}$ is chosen to be blocked self-consistently. It shows that when calculations are performed without blocking, both proton and neutron density distributions of $^{17}$B have a spherical shape, in both SLy4 and UNEDF1 calculations. However, when calculations are performed with the blocking of proton $1p_{1/2}$,
 both proton and neutron density distributions of $^{17}$B have obvious prolate deformation in the inner part and in the surface. The onset of deformation is due to the polarization effect caused by blocking. The total quadrupole deformation of $^{17}$B with blocking is $\beta_2=0.30$ with SLy4. The relativistic Hartree-Bogoliubov model calculations of $^{17}$B without blocking also obtained large deformations\cite{Lalazissis04}.  In our calculations, the halo structure in $^{17}$B is not so impressive. Experimentally, the density distributions of $^{17}$B  can be extracted but are rather model dependent.
 The density profiles of $^{19}$B are displayed in Fig.\ref{figB19}. Compared to $^{17}$B,  one can see that $^{19}$B has a more pronounced halo structure with or without blocking calculations using SLy4 and UNEDF1 parameters. Similar to $^{17}$B, $^{19}$B has a prolate deformed inner part with blocking. Its halo is prolately deformed with SLy4 and is spherical with UNEDF1.

Fig.\ref{figMg37} displays the density profiles of $^{37}$Mg. $^{37}$Mg has no obvious halo without blocking and shows a clear deformed halo with blocking calculations.
Recent experimental results demonstrated that the last bound neutron level of $^{37}$Mg has an important (about $40\%$) $p$-wave component\cite{Mg37}. Another most likely component of the level is the $f$-wave which is close to the $p$-wave. A high centrifugal barrier of $f$-wave would suppress the halo formation.  Note that in the deformed case we are blocking the $\Omega^{\pi}$=1/2$^{-}$ state that naturally contains both components. The remarkable halo structure indicated the role of $p$-wave.   We see that blocking the neutron orbit has little effect on proton densities of $^{37}$Mg. Deformation of the inner part in $^{37}$Mg has little change with and without blocking, in contrast to $^{17,19}$B.

Fig.\ref{figMg37p} shows the pairing density profiles of $^{37}$Mg. In weakly-bound even-even nuclei, the pairing density distributions generally have more pronounced halo structures than
the particle density distributions. This is also shown in $^{37}$Mg without blocking. However, the pairing density in $^{37}$Mg does not show a halo feature when blocking is included.
Indeed, the pairing correlations have been significantly reduced due to the blocking effect.
Along the $z$-axis, the pairing density decreases rapidly and goes to negative values after 17 fm,  emerging small amplitude oscillations. This has an analogy to the Larkin-Ovchinnikov (LO) phase in spin-polarized cold Fermi gases in elongated traps~\cite{Pei10}, although the spin polarization in $^{37}$Mg is very small ($\simeq\frac{1}{25}$).

Table \ref{table1}. lists the calculation results of $^{17}$B, $^{19}$B and $^{37}$Mg. In general, the blocking effect results in less binding energies and larger deformations.
 It shows that $^{17}$B has less obvious halo structure in calculations with SLy4 parameters and UNEDF1 parameters whether blocking is included or not. Experimental results\cite{Suzuki99} show the last two neutrons of $^{17}$B have quite a few $d$-wave component which would hinder halo formation. The last unpaired proton of boron occupies the $p$ orbit with and without blocking.
The calculated proton r.m.s radii are almost unchanged in both approaches. However, the neutron r.m.s radii reduce slightly with blocking, as shown in Table \ref{table1}.
  The latest experimental value of r.m.s mass radius of $^{17}$B is $3.00\pm0.06$ fm\cite{Estrade14}. The closest result of present work is $3.06$ fm, from the blocking calculation with SLy4 force. For $^{19}$B, the experimental r.m.s radius is $3.11\pm0.13$ fm, which is smaller than our theoretical results. The two-neutron separation energy
  of $^{19}$B is 0.23 MeV with SLy4 blocking calculations while the experimental value is 0.14$\pm0.39$ MeV\cite{Gaudefroy12}.
   The total deformation increases and the absolute value of total binding energy reduces in $^{37}$Mg after the blocking. The large deformations in $^{37}$Mg are mainly due to the contributions from large spatial extensions.   Note that light weakly bound nuclei have very soft deformations. The one-neutron separation energy
  of $^{37}$Mg is 0.43 MeV with UNEDF1 blocking calculations while the experimental value is 0.22$_{-0.09}^{+0.12}$ MeV\cite{Mg37}. One can see that theoretical descriptions of drip line nuclei are strongly dependent on the effective nuclear interactions.

\section{Conclusion}

 In summary,  the Hartree-Fock-Bogoliubov approach with self-consistent quasiparticle blocking in deformed coordinate space has been applied to study the neutron-rich odd-A halo nuclei $^{17}$B, $^{19}$B and $^{37}$Mg. This is the first time such blocking calculations of odd-A nuclei in deformed coordinate spaces has been done, and it is an ideal tool for describing deformed halos and continuum effects together. To check the dependence of the parameters, two Skyrme forces, SLy4 and UNEDF1, are adopted. Generally, the odd-A nuclei become less
 bound and deformations increase due to blocking effects. In particular, a remarkable deformed halo structure in $^{37}$Mg has been obtained by blocking the $\Omega^{\pi}$=1/2$^{-}$ neutron state, and it is consistent with a large $p$-wave component. However, $^{17}$B has less obvious halos, regardless of whether or not blocking is used in the calculation, compared to $^{19}$B. The indication of Larkin-Ovchinnikov  phase in $^{37}$Mg has also been discussed.

\acknowledgments{Discussion with Xu F R, Peking University, is gratefully acknowledged.}

\end{multicols}

\vspace{10mm}

\vspace{-1mm}
\centerline{\rule{80mm}{0.1pt}}
\vspace{2mm}

\begin{multicols}{2}

\end{multicols}

\clearpage
\end{CJK*}

\begin{thebibliography}{90}
\vspace{3mm}


\bibitem{halo-exp} TANIHATA I. J. Phys. G, 1996, {\bf 22}: 157

\bibitem{FRIB} http://www.frib.msu.edu

\bibitem{HIAF} YANG J C et al. Nucl. Instr. Meth. B, 2013, {\bf 317}: 263---265

\bibitem{Pei13} PEI J C, ZHANG Y N, XU F R. Phys. Rev. C, 2013, {\bf87}: 051302(R)

\bibitem{Pei14a} PEI J C, KORTELAINEN M, ZHANG Y N, XU F R. Phys. Rev. C, 2014, {\bf 90}: 051304


\bibitem{RBF1} Li L L, Meng J, Ring P, Zhao E G, Zhou S G. Phys. Rev. C, 2012, {\bf 85}: 024312


\bibitem{sgzhou}
ZHOU S G, MENG J, RING P, ZHAO E G. Phys. Rev. C, 2010, {\bf 82}: 011301


\bibitem{Be11} FUKUDA M et al. Phys. Lett. B, 1991, {\bf 268}: 339---344

\bibitem{Ne31} NAKAMURA T et al. Phys. Rev. Lett., 2009, \textbf{103}: 262501

\bibitem{Mg37} KOBAYASHI N et al. Phys. Rev. Lett., 2014 \textbf{112}: 242501

\bibitem{Bertsch} BERTSCH G, DOBACZEWSKI J, NAZAREWICZ W, PEI J C. Phys. Rev. A, 2009, {\bf 79}: 043602

\bibitem{Suzuki99} SUZUKI T et al. Nucl. Phys. A, 1999, {\bf 658}: 313---326

\bibitem{Suzuki02} SUZUKI T et al. Phys. Rev. Lett., 2002, {\bf 89}: 012501

\bibitem{Ren} REN Z Z, XU G O. Phys. Lett. B, 1990, {\bf 252}: 311---313

\bibitem{HU08} HU Z G et al. Sci. China Ser. G-Phys. Mech. Astron., 2008, {\bf 51}: 781---787

\bibitem{xiajijuan12} JI J X, LI J X, HAN R, WANG J S, HU Q. Chin. Phys. C, 2012, {\bf 36}: 43---47

\bibitem{Guoyanqing10} GUO Y Q, REN Z Z. Chin. Phys. Lett., 2010, {\bf 27}: 102102

\bibitem{WangM08} WANG M et al. Chin. Phys. C, 2008, {\bf 32}: 548---551

\bibitem{Gaudefroy12} GAUDEFROY L et al. Phys. Rev. Lett., 2012, {\bf109}: 202503

\bibitem{Mazy}
LIANG J, CAO L G, MA Z Y. Phys. Rev. C, 2007, {\bf 75}: 054320

\bibitem{Pei08} PEI J C, STOITSOV M V, FANN G I, NAZAREWICZ W, SCHUNCK N, XU F R. Phys. Rev. C, 2008, {\bf 78}: 064306



\bibitem{Pei14} PEI J C, FANN G I, HARRISON R J, NAZAREWICZ W, SHI Y, THORNTON S. Phys. Rev. C, 2014, {\bf 90}: 024317

\bibitem{sly4}
 CHABANAT E, BONCHE P, HAENSEL P, MEYER J, SCHAEFFER R. Nucl. Phys. A, 1998, {\bf 635}: 231--256


\bibitem{unedf1}
KORTELAINEN M, MCDONNELL J, NAZAREWICZ W, REINHARD P -G, SARICH J, SCHUNCK N, STOITSOV M V, WILD S M. Phys. Rev. C, 2012, {\bf 85}: 024304

\bibitem{cheny}
CHEN Y, RING P, MENG J. Phys. Rev. C, 2014, {\bf 89}: 014312

\bibitem{Cwiok99} \'{C}WIOK S, NAZAREWICZ W, HEENEN P  H. Phys. Rev. Lett., 1999, {\bf 83}: 1108---1111

\bibitem{Xu98} XU F R, WALKER P M, SHEIKH J A, WYSS R. Phys. Lett. B, 1998, {\bf 435}: 257---263

\bibitem{RHB2} Li L L, Meng J, Ring P, Zhao E G, Zhou S G. Chin. Phys. Lett, 2012, {\bf 29}: 042101


\bibitem{Pei10}
PEI J C, DUKELSKY J, NAZAREWICZ W. Phys. Rev. A, 2010, {\bf 82}: 021603(R)

\bibitem{2FLA}
SENSARMA R, SCHNEIDER W, DIENER R B, RANDERIA. arXiv, 2007, 0706.1741v1.

\bibitem{Lalazissis04} LALAZISSIS G A, VRETENAR D, RING P. Eur. Phys. J. A, 2004, {\bf 22}: 37---45

\bibitem{Estrade14} ESTRAD\'{e} A et al. Phys. Rev. Lett., 2014, {\bf 113}: 132501



























\end{thebibliography}
\end{document}